\documentclass[12pt]{article}
\usepackage[utf8]{inputenc}
\usepackage[english]{babel} 
\usepackage[dvips,pdftex]{graphicx}
\usepackage[round]{natbib}
\usepackage{hyperref}
\graphicspath{{images/}}
\newcommand{\angstrom}{\textup{\AA}}
\newcommand{\footremember}[2]{%
    \footnote{#2}
    \newcounter{#1}
    \setcounter{#1}{\value{footnote}}%
}
\title{Masses of optical components and black holes in x-ray novae: the effects of components proximity.}
\author{Petrov $^1$\footremember{alley}{patrokl@gmail.com}V. S., Antokhina$^1$ E. A.,  Cherepashchuk$^1$ A. M.}
\date{%
    $^1$Sternberg Astronomical Institute, Moscow M.V. Lomonosov State University\\[2ex]%
    \today}
\begin{document}
\maketitle
\begin{abstract}
An exact calculation of CaI $\lambda 6439.075 \; \angstrom $ absorption profiles in the spectra of optical stars in low-mass X-ray binary systems with black holes (BH LMXBs) is carried out. We show that the approximation of a real Roche lobe filling star as disk with uniform local line profile and linear limb darkening law leads to overestimation of  projected equatorial rotational velocity $V_{rot} \sin i$ and accordingly, underestimation of mass ratio $q=M_x/M_v$. Refined value of $q$ does not affect the mass of a black hole, but the mass of an optical star has shrunk $\sim 1.5$ times. We present refined components masses in low-mass X-ray binaries with black holes. Companion masses in BH LMXBs are found in the mass range 0.1 - 1.6 $M_{\odot}$ with the peak at $M_v \simeq 0.35 M_{\odot}$. This finding poses additional problem for the standard evolutionary scenarios of BH LMXBs formation with a common envelope phase (CE). We also discuss the implications of these masses into the evolutionary history of the binary.
\end{abstract}

\section{Introduction}
The mass ratio $q=M_x/M_v$ is an fundamental parameter of the evolution of the low-mass X-ray binary system. The binary mass ratio $q=M_x/M_v$ ($M_x$ is the mass of the black hole and $M_v$ is the mass of optical star) is obtained by measuring the rotational velocity of the secondary star $V_{rot} \sin i$ \citep{Wade_Horne_1988}. The procedure commonly used to measure of the low-mass X-ray binaries (LMXBs) secondary star rotational broadening is to convolve its spectrum with a limb-darkened standard rotation profile \citep{Collins_1995}. It used the observed line profiles for slowly rotating stars with close spectral types (reference stars) as profiles unbroadened by rotation. The spectra of the stars in the X-ray binaries and those of the single slowly rotating stars were obtained with the same spectral resolution. The rotational broadening of the spectra of the reference stars was modelled, assuming that these profiles would be the same in the absence of rotation, identifying the value of $V_{rot} \sin i$ for which the spectra of the single star and the star in the X-ray binary were in best agreement with $\chi^2$ criterion. In this way, the rotational velocity $V_{rot} \sin i$ of the star in the binary system was found. This approach made it possible not to correct the influence of the instrumental function of the spectrograph. 
In the up to date catalogue of black hole transients BlackCat \citep{Casares_Jonker_2014} authors shows the data about the rotational broadening of line profiles in the spectra of the optical components of X-ray binary systems/ obtaining by using classical rotational broadening model \citep{Collins_1995}.

An exact calculation of CaI $\lambda 6439.075 \; \angstrom $  absorption profiles in the spectra of optical stars in low mass X-ray binary systems is carried out \citet{Petrov_2017}. They showed that the Full Width Half Maximum ($FWHM$) of CaI $\lambda 6439.075 \; \angstrom $  absorption profiles in tidally deformed Roche model  \citep{Antokhina_Shim_2005} is completely higher than the $FWHM$ of CaI $\lambda 6439.075 \; \angstrom $  absorption profiles in classical rotational broadening model \citep{Collins_1995}. It was shown that widely used approximation of optical star in LMXB as a disk with an uniform local profile and linear limb darkening law leads to overestimation of  projected equatorial rotational velocity $V_{rot} \sin i$ and, accordingly, underestimation of mass ratio $q=M_x/M_v$. \citet{Petrov_2017} presented an equation that allows to convert the mass ratio $q_{disk}$ obtained  with the classical rotational broadening model to the mass ratio $q=M_x/M_v$ corresponding to Roche model \citep{Antokhina_Shim_2005}. 

There is a mismatch between modelled and observed distributions of optical stars and black holes masses in BH LMXBs (e.g. \citep{Podsiadlowski_2010}). For example, the BH mass spectrum is flat, with the gap in the mass range $2–5 M_{\odot}$ \citep{Ozel_2010, Farr_2011, Petrov_2014}. Companion masses in BH LMXB are found in the mass range 0.1 - 1.6 $M_{\odot}$ with the peak at 0.6 $M_{\odot}$. It seems rather difficult to explain the existence of such systems with the standard common envelope evolutionary scenarios: ejection of massive common envelope of such low-mass star proves rather difficult.\citet{Wang_2016} proposed that it is possible to form BH LMXBs with the standard CE scenario if most BHs  are born through failed supernovae with  progenitor  mass $ M<28 M_{\odot} $. But $\alpha$ - elements (O, Si, Mg e.t.c.) are found in atmospheres of at least two optical counterparts: GRO J1655-40 and SAX J1819.3-2525 \citep{Israelian_1999}, so found evidence of the formation of the black holes in an explosive supernova (SN) event. 
In this paper we refined mass ratio $q$ for 9 BH LMXBs listed in BlackCAT catalogue \citep{Casares_Jonker_2014} using the approximation equation from Antokhina et al. \citep{Petrov_2017}. We assume that approximation equation based on CaI $\lambda 6439.075 \; \angstrom $  absorption line profiles can also be applied to other lines that are not subject to strong Stark effects.  The radial velocity curves of the optical stars in these systems differ from the radial velocity curves of their barycenters, due to the effects of tidal distortion, gravitational darkening, X-ray heating, etc. Hence, the mass functions $f_v(M)$ derived from these radial velocity curves will differ accordingly. We took into account the closeness of the components when determining masses of components using the K-correction method \citep{Petrov_2013}. Refined value of $q$ does not affect the mass of a black hole, but the mass of an optical star in most cases decreased up $\sim 1.5$ times. 

\section{Parameters of low mass X-ray binary systems containig black holes}

The physical parameters of 9 BH LMXBs are listed in Table \ref{tabular:parameters}. The spectra of optical counterparts received with intermediate resolution spectroscopy. This spectra allow to estimate rotational broadening of absorption lines, the radial velocity semi amplitudes, and the orbital inclination combined with a light curves. 

In the Table \ref{tabular:parameters}  $V_{rot} \sin i$  is the projected equatorial rotational velocity, $K_v$ is the observed radial velocity semi-amplitude, $f_{v}(M)$ is the observed mass function, $i$ is the orbital inclination (see \citep{Lutiy_1973}).

Note, \citet{Shahbaz2014} showed, that using relatively low-resolution spectroscopy can result in systemic uncertainties in the measured $V_{rot} \sin i$ values obtained classical rotational broadening method.
The mass ratio $q=M_x/M_v$ we estimate with the following equation \citep{Wade_Horne_1988}:
\begin{eqnarray} \label{eq:Pach}
\frac{V_{rot}\sin i}{K_{c}}\simeq 0.462\; q^{-1/3}\left( 1+\frac{1}{q}\right)^{2/3}.
\end{eqnarray}

It is necessary to mention that the radial velocity semi-amplitude of the optical stars $K_v$ are listed in Table \ref{tabular:parameters} differ from the the radial velocity semi-amplitude of their barycenters $K_c$, due to the effects of tidal distortion, gravitational darkening, X-ray heating, etc. We took into account the closeness of the components using the K-correction method \citep{Wade_Horne_1988, Petrov_2013}.
The mass of black hole $M_x$ was determined using the equation:
\begin{eqnarray} \label{eq:massfun}
M_{x} &=& \frac{f_{v}(M)\left(1+q^{-1}\right)^{2}}{\sin^{3} i},
\end{eqnarray} 
and the optical star mass is:
$$
M_v=M_x/q.
$$
Here $f_{v}(M) = 1.038 \cdot 10^{-7} K_v^{3} P_{orb}(1-e^2)^{3/2}$ is mass function, $K_v$ is observed radial velocity semi-amplitude of the optical stars, $e$ is the orbital eccentricity (for BH LMXBs $e=0$),  $P_{orb}$ is the orbital period of the binary.

In the limit case where $q=M_x/M_v >> 1$ (this case corresponds to BH LMXB) mass ratio $q$ is given by:
\begin{eqnarray} \label{eq:big_q}
q \simeq \left( \frac{0.462 K_c}{V_{rot} \sin i } \right)^{3}.
\end{eqnarray}
It follows from the expression (\ref{eq:big_q})  that small uncertainties in $V_{rot} \sin i$  lead to large uncertainties in $q$.
\section{Refined black hole and optical star masses distribution}
We use a method of mass ratio correction developed by \citet{Petrov_2017} to study the black hole and optical star masses. As mentioned above we assume that approximation equation based on CaI $\lambda 6439.075 \; \angstrom $  absorption line profiles can also be applied to other lines that are not subject to strong Stark effects. For BH LMXB in quiescent the effect of X-ray heating is ignored: $k_x=L_x/L_{v}^{bol}=0$ (the effect of heating by the X-ray radiation is described by \citet{Petrov2015}).

The  method of mass ratio correction developed by \citet{Petrov_2017} provides a corrected value of the binary mass ratio through the expression:
\begin{eqnarray}\label{eq:NonLinFitq}
q_{corr} = q_{disk} + \Delta q,
\end{eqnarray} 
where the $\Delta q$ correction is described by
\begin{eqnarray}\label{eq:NonLinFitDeltaq}
\Delta q = ( 0.41 \pm 0.01) \cdot {q_{disk}} ^{1.224 \pm 0.008}.
\end{eqnarray} 

\citet{Petrov_2017} showed that equation (\ref{eq:NonLinFitq}) is accurate better than  $5 \; \%$ in effective temperature range $ 4000\; K$ - $8000 \; K$.  
We also took into account the closeness of the components when determining $M_x$. Hence, the mass functions $f_v(M)$ derived from observed radial velocity semi-amplitude of the optical stars will be changed with the following equation:
\begin{eqnarray} \label{eq:defnussfunc}
f_v^{corr}(M)=1.038\cdot 10^{-7} K_c^3P_{orb},
\end{eqnarray}
where $K_c$ is the semi-amplitude of the radial velocity curve of the stellar barycenter. The $K_c$ is given by $K_c = K_v/K_{corr}(q,i)$ where $K_v$ is observed radial velocity semi-amplitude of the optical stars and $K_{corr}$ is the K-correction. For each system in Table \ref{tabular:parameters}, the corresponding K-correction was chosen from \citet{Petrov_2013,Petrov2015}.
In next step, we corrected masses of optical stars: $M_v=M_x/q$.

\begin{figure}[h]
\center{\includegraphics[width=1\linewidth]{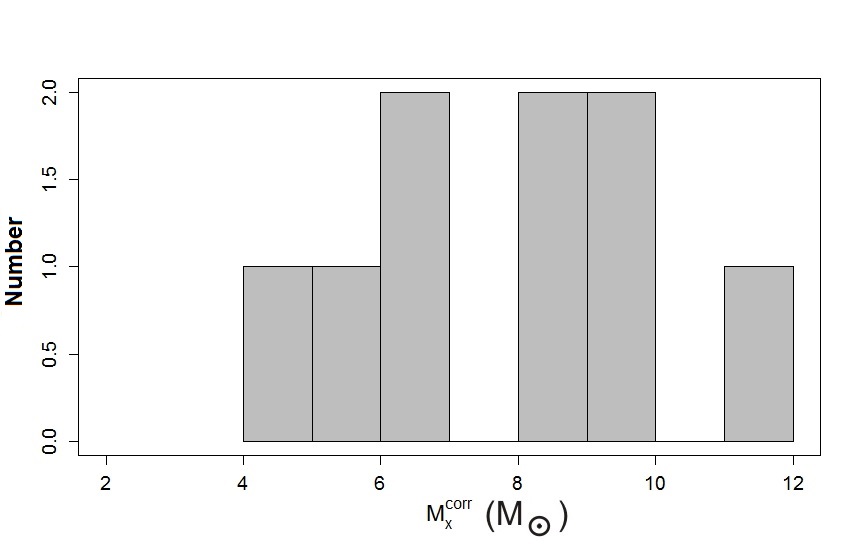}}\\
\caption{The refined mass distribution of black holes (see table. \ref{tabular:parameters})}
\label{ris:Mx_hist}
\end{figure}

\begin{figure}[h]
\center{\includegraphics[width=1\linewidth]{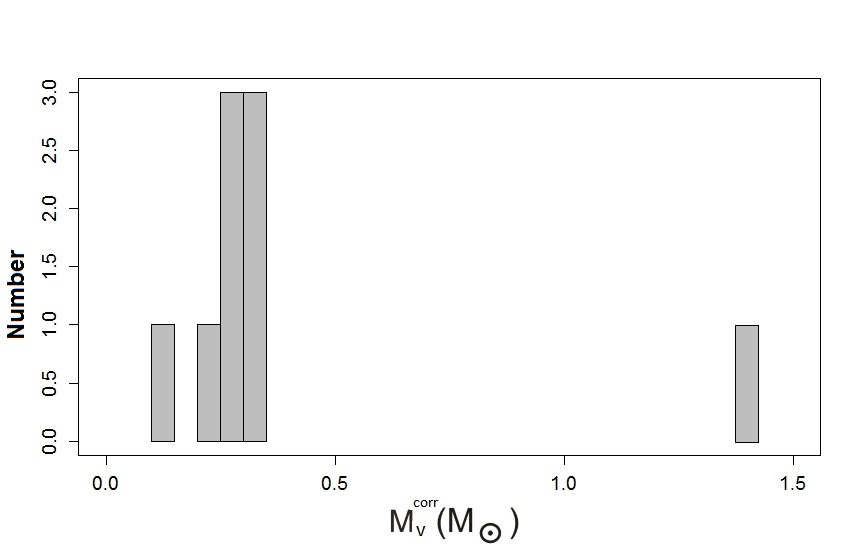}}\\
\caption{The refined mass distribution of optical counterparts in BH LMXBs (see table. \ref{tabular:parameters})}
\label{ris:Mv_cor_tot}
\end{figure}

Initial and refined masses of black holes and optical stars are listed in Table \ref{tabular:Corparameters}. We also present the refined distribution of the black hole masses in Fig. \ref{ris:Mx_hist} and the refined distribution of the optical star masses in Fig \ref{ris:Mv_cor_tot}.
Refined value of $q$ does not affect the mass of a black hole, but  the mass of an optical star has in the most cases shrunk $\sim 1.5$ times (see Table \ref{tabular:Corparameters}). The system GRO 1655-40 deserve special mention. The corrected black hole mass $M_x$ has the bias due the K-correction using.  

\section{Discussion}

We present the refined distribution of the donor masses in BH LMXBs with the peak at 0.35 $M_{\odot}$ (excluding the system GRO 1655-40). This distribution not only allow to estimate realistic binding energy of the common envelope $\lambda$, but also to bring several alternatives to the observations.


One of the alternative evolution scenario is that the secondaries in BH LMXBs enhanced the mass loss rate due to illumination of the stellar surface by the high energy radiation from the accretion disk around the BH during outburst \citep{Arons_1973,Basko_1973,Basko_1974,Basko_1977}. This scenario produced a similar companion masses distribution with peak at $\sim 0.4 \;M_{\odot}$ \citep{Wiktorowicz_2014}. 

In the irradiation-induced wind scenario the mass transfer is non conservative \citep{ITY1995}. The black hole in an LMXB typically accretes only about 10\% of the mass lost by the donor. A coronal wind carries about 90\% mass lost by the donor away. Thus the increase of black hole mass should be negligible. Apparently, coronal wind velocity not enough to blow out the mass of the system and matter can form a disk-shaped circumbinary envelope.

Spectroscopic observations confirm fast orbital decays of black hole X-ray binaries: XTE J1118+480 ($\dot{P}=-1.90\pm 0.57$ ms/yr) and A0620-00 ($\dot{P}=-0.60\pm 0.08$ ms/yr) \citep{Gonzales_2014}. Angular momentum losses due to gravitational radiation and magnetic braking are unable to explain these large orbital decays in these two short-period black hole binaries. The fast spiral-in of the star in BH LMXB does not fit the rapid evaporation of stellar-mass black holes \citep{Postnov_2003} in multi-dimensional models of gravity \citep{Emparan_2003} on the RS brane \citep{Randall_1999}. 
\citet{Chen_2015}  showed that, for some reasonable parameters, tidal torque between the circumbinary disk and the binary can efficiently extract the orbital angular momentum from the binary. Observations have provided evidence that circumbinary disks around two compact black hole X-ray binaries may exist. \citet{Muno_2006}  have detected the blackbody spectrum of  BHXBs A0620-00 and XTE J1118+480, and found that the inferred areas of mid-infrared 4.5-8 $\mu m$ excess emission are about two times larger than the binary orbital separations. A detection with the Wide-field Infrared Survey Explorer identified that XTE J1118+480 and A0620-00 are candidate circumbinary disk systems \citep{Wang_2014}.
 
The extremely low-mass optical counterparts of BH LMXBs ( with masses $M_v < 0.35 \; M_{\odot}$) can result of evolution massive close binary accompanying at large distance by a G-K main sequence dwarf \citep{Eggleton_Verbunt_1986}. After the evolution of the close binary into an ordinary X-ray binary, the compact object is engulfed by its expanding massive companion, and spirals in to settle at its centre. The resulting Thorne-Zytkow supergiant gradually expands until it attains the size of the main sequence dwarf  orbit. Then a second spiral-in phase ensues, leading to the formation of a low-mass close binary. 
Tidal capture formation scenario, also can correspond  to this refined distribution \citep{Erez_2016}. In this formation channel, LMXBs are formed from wide binaries ($>1000$ au) with a BH component and a stellar companion. Resulting optical mass distribution is this scenario peaking at $0.4-0.6 \; M_{\odot}$. Note, that in this case the LMXBs orbit should not correlate with the spin of the BH. 
\section{Conclusions}
We present refined binary masses in low-mass X-ray binaries with black holes. The refined masses of an optical stars have shrunk $\sim 1.5$ times. In this way the refined distribution of the donor masses in BH LMXBs has the peak at 0.35 $M_{\odot}$ ($\bar{M}_v=0.31 \; M_{\odot}$ without GRO 1655-40). This distribution is well described by irradiation-induced wind evolution scenario.  Furthermore, The extremely low-mass optical counterparts of BH LMXBs ($M_v<0.35 \; M_{\odot}$) can result of evolution massive close binary accompanying at large distance by a G-K main sequence dwarf \citep{Eggleton_Verbunt_1986} or tidal capture formation scenario \citep{Erez_2016}.

We thank Dmitry V. Bisikalo, Konstantin A. Postnov and Chris Belczynski  for their useful comments. 

\bibliographystyle{apj}
\bibliography{main}

\newpage
\begin{table}[h]
\footnotesize
\caption{The physical parameters of 9 BH LMXB with surely estimated the rotational broadening
of the secondary star  $V_{rot} \sin i$ from BlackCAT catalogue \citep{Casares_Jonker_2014}}   
\label{tabular:parameters}
\begin{center}
\setlength{\tabcolsep}{5pt}
\begin{tabular}{lcccccc}
\hline
\hline
System &$V_{rot} \sin i$, km/s &$K_v$, km/s &$f_v(M),\;M_{\odot}$ &$q=M_x/M_v$& $i$, deg.& Links \\
\hline
A0620-00 &$83 \pm 5$&$437 \pm 2$&$2.80\pm 0.01$&$16 \pm 3$&$51 \pm 1$ 
&1,2\\
GS 2023+338&$39.1 \pm 1.2$&$208.5\pm 0.7$&$6.08\pm 0.06$&$16.7 \pm 1.4$&$ 66-70 $
&3\\
GS 2000+251&$86 \pm 8$&$520 \pm 5$&$5.01\pm 0.12$&$24 \pm 7$&$ 43-74 $
&4\\ 
GRO 1655-40&$88\pm 5$&$226.1\pm 0.8$&$2.73\pm 0.09$&$3.0 \pm 0.3$&$70\pm 2$
&5\\ 
XTE J1118+48&$96^{+4}_{-11} $&$709 \pm 1$&$6.28\pm 0.04$&$41^{+5}_{-16}$&$68\pm 2$
&6\\ 
GRS 1915+105&$21 \pm 4 $&$126\pm 1$&$7.02 \pm 0.17$&$23 \pm 2$&$66\pm 2$
&7 \\ 
GRS 1009-45&$ 86 \pm 5$&$475\pm 6$&$3.17\pm 0.12$&$18 \pm 3$&$37-80$
&8 \\
GRO J0422+32&$ 90 \pm 22$&$378\pm 16$&$1.19\pm 0.02$&$11 \pm 6$&$45\pm 2$
&9\\ 
GRS 1124-68&$ 106 \pm 13$&$406 \pm 7$&$3.01\pm 0.15$&$13 \pm 1$&$54\pm 2$
&10\\ 
\hline
\hline
\multicolumn{7}{l}{\rule{0pt}{11pt}\footnotesize  1- \citep{Gonzales_2014};2-\citep{Marsh1994};3-\citep{Casares1994}; }\\
\multicolumn{7}{l}{\rule{0pt}{11pt}\footnotesize  4-\citep{Harlaftis1996}; 5-\citep{Shahbaz2003};6-\citep{Calvelo2009}; 7-\citep{Steeghs2013}  }\\
\multicolumn{7}{l}{\rule{0pt}{11pt}\footnotesize  8- \citep{Filippenko1999}; 9-\citep{Webb2000}; 10-\citep{Orosz1996}  }\\
\multicolumn{7}{l}{\rule{0pt}{11pt}\footnotesize  Note: the observed radial velocity semi-amplitude $K_v$ which allow to calculate observed }\\
\multicolumn{7}{l}{\rule{0pt}{11pt}\footnotesize  mass function $f_v(M)$; The classical rotational broadening model was used for }\\
\multicolumn{7}{l}{\rule{0pt}{11pt}\footnotesize $V_{rot} \sin i$ determination }\\ 
\end{tabular}
\end{center}
\end{table}
%
%
\begin{table}[h]
\footnotesize
\caption{Refined parameters of BH LMXBs}
\label{tabular:Corparameters}
\begin{center}
\setlength{\tabcolsep}{3pt}
\begin{tabular}{lccccccc}
\hline
\hline
 System&$q_{corr}$&$K_c$, km/s &$f_v^{corr}(M),M_{\odot}$ &$M_x$, $ M_{\odot} $&$M_x^{corr}$, $ M_{\odot} $&$M_v$, $ M_{\odot} $&$M_v^{corr}$, $ M_{\odot} $\\
 \hline
A0620-00 & $26 \pm 5$&$438.4 \pm 2.0$& $2.82\pm 0.036$& $6.7\pm 0.3 $ &  $6.5 \pm 0.3$ & $0.4 \pm 0.1 $&$ 0.25 \pm 0.1 $ \\
GS 2023+338& $26 \pm 2 $& $209.1\pm 0.7$& $6.14\pm 0.06$ & $8.2 -8.9$ &  $7.9 - 8.6$&$0.5\pm 0.1 $& $0.3 \pm 0.1 $ \\
GS 2000+251& $39 \pm 8$& $521\pm 5$& $5.06\pm 0.12$ & $6.1 - 17.1 $ &  $6.0-16.7$ &$0.2 \div 0.7 $&$0.1  \div  0.4$\\ 
GRO 1655-40& $3.9\pm 0.5$& $229.6\pm 2.4$& $3.29\pm 0.09$&$5.8 \pm 0.4 $&$6.0 \pm 0.4$ & $ 1.9 \pm 0.2$ & $ 1.4 \pm 0.2 $\\ 
XTE J1118+48&  $73 ^{+10}_{-32} $&  $710.2\pm 1.4$&  $6.3 \pm 0.1$ &$8.2 \pm 0.4 $&$8.1 \pm 0.4$& $0.20 \pm 0.04$ &$ 0.10 \pm 0.04 $\\ 
GRS 1915+105&$35 \pm 13 $&$126\pm 1$&$7.07\pm 0.17$&$10.1 \pm 0.6$&$9.8 \pm 0.6$&$0.4 \pm 0.1$ &$ 0.3 \pm 0.1$\\ 
GRS 1009-45&$30 \pm 5$& $476\pm 4$&$3.20\pm 0.12$&$3.7  - 16.2$&$3.6-15.7$&$0.2 \div 0.9$ &$0.1  \div 0.5 $\\
GRO J0422+32&$ 14 \pm 3$&$379\pm 16$&$1.20\pm 0.02$&$4.1 \pm 0.6$&$3.9 \pm 0.6$&$0.5 \pm 0.1$&$ 0.3 \pm 0.1$\\ 
GRS 1124-68&$ 21 \pm 1$&  $407.6 \pm 2.0$&  $3.04 \pm 0.15$& $6.6 \pm 0.5$ &  $6.3 \pm 0.5$&$ 0.5 \pm 0.1$&$ 0.3 \pm 0.1$ \\ 
%
%
%
\hline
\hline
\multicolumn{8}{l}{\rule{0pt}{11pt}\footnotesize  Note: $M_x$ is the black hole mass, $M_v$ is the optical star mass defined with the classical }\\
\multicolumn{8}{l}{\rule{0pt}{11pt}\footnotesize rotational broadening. }\\
\end{tabular}
\end{center}\end{table}

\end{document}